\documentclass[journal]{IEEEtran}
\IEEEoverridecommandlockouts                              % This command is only
\usepackage{graphicx} % for pdf, bitmapped graphics files
\usepackage{amsmath,amssymb, amsfonts, amsthm}
\usepackage{amsmath} % assumes amsmath package installed
\usepackage{comment} 
\usepackage{algorithm,algcompatible}
\usepackage{color}
\usepackage[english]{babel}
\usepackage[noend]{algpseudocode}
\usepackage{mathrsfs}
\usepackage[justification=centering]{caption}

\newtheorem{lemma}{Lemma}
\newtheorem{theorem}{Theorem}
\newtheorem{problem}{Problem}

\newtheorem{definition}{Definition}
\newtheorem{remark}{Remark}
\newtheorem{example}{Example}

\title{\LARGE \bf
Supervisor Obfuscation Against Actuator Enablement Attack
} 
\author{Yuting Zhu*, Liyong Lin*, Rong Su
\thanks{Y.~Zhu, L.~Lin and R.~Su are affiliated with School of Electrical and Electronic Engineering, Nanyang Technological University, Singapore. Email:  yuting002@e.ntu.edu.sg, liyong.lin@utoronto.ca,  rsu@ntu.edu.sg. This work is financially supported by Singapore Ministry of Education Academic Research Grant RG91/18-(S)-SU RONG (VP), which is gratefully acknowledged.

*Yuting Zhu and Liyong Lin contribute equally to this work. Liyong Lin would like to acknowledge the discussion with Prof. Wonham on the SAT based approach for the synthesis of behavior-preserving supervisors during 2016 and 2017.
}%
}
\begin{document}
\maketitle
\thispagestyle{empty}
\pagestyle{empty}

%%%%%%%%%%%%%%%%%%%%%%%%%%%%%%%%%%%%%%%%%%%%%%%%%%%%%%%%%%%%%%%%%%%%%%%%%%%%%%%%
\begin{abstract}
In this paper, we propose and address the problem of supervisor obfuscation against actuator enablement attack, in a common setting where the actuator attacker can eavesdrop the control commands issued by the supervisor. We propose a method to obfuscate an (insecure) supervisor to make it resilient against actuator enablement attack in such a way that the behavior of the original closed-loop system is preserved. An additional feature of the obfuscated  supervisor, if it exists, is that it has exactly the minimum number of states among the set of all the resilient and behavior-preserving supervisors. Our approach involves a simple  combination of two basic ideas: 1) a formulation of the problem of computing behavior-preserving supervisors  as the problem of computing separating finite state automata under controllability and observability constraints, which can be efficiently  tackled by  using modern SAT solvers, and 2) the use of a recently proposed technique for the verification of attackability in our setting, with a normality assumption imposed on both the actuator attackers and supervisors. %A case study is also carried to illustrate the effectiveness of our approach.    
%The problem of supervisor reduction is known to be NP-complete. In this work, we show that  supervisor reduction can be viewed as a simple generalization of the problem of computing minimal separating DFAs (deterministic finite automata). Based on this observation, we introduce some different techniques that can be used to compute minimal reduced supervisors. {\color{red} XXX}%We then conduct experiments and compare these solutions with the supervisor reduction heuristic available in the literature. 
%those of heuristic solutions available in the literature. %, which is a well studied NP-complete problem and has many exact or approximated solutions  in the formal verification and automaton learning research communities , which is a well studied NP-complete problem
\end{abstract}
\vspace{-4pt}
\section{Introduction}
Recently, cyber-physical systems have drawn a lot of attention from the supervisory control and formal methods research communities (see, for example,~\cite{CarvalhoEnablementAttacks},~\cite{Su2018},~\cite{Goes2017},~\cite{Lanotte2017},~\cite{Jones2014},~\cite{WTH17},~\cite{LACM17}). The supervisory control theory of discrete-event systems\footnote{In this work, we focus on the class of cyber-physical systems that can be modelled using discrete-event systems.}~\cite{WMW10} has been proposed as a general approach for the synthesis of  correct-by-construction supervisors that ensure both safety and progress properties on the closed-loop systems. However, the correctness guarantee is implicitly based upon the assumption that the system has been fully and correctly modelled. In the presence of unmodeled adversarial attackers, the synthesized supervisor cannot guarantee safety any more. Thus, one is then required to synthesize resilient supervisors against adversarial attacks. %, which is reduced to constructively proving the following.
%\begin{center}
%$\exists S  \forall A, \Phi_{(G, S, A, H)}^{safe}$,
%\end{center}
%That is, there exists a supervisor 
In view of the importance and prevalence of cyber-physical systems in modern society~\cite{RLSS10, MKBDLP12, D13},  researchers in supervisory control theory have suggested various frameworks and techniques (see~\cite{CarvalhoEnablementAttacks},~\cite{Su2018},~\cite{WTH17},~\cite{LACM17}) for the design of mitigation and resilient control mechanisms against adversarial attacks.
 In this work, we follow the setup of~\cite{Lin2018} and consider the problem of synthesis of resilient supervisors against actuator enablement attack, with a normality assumption imposed on both the actuator attackers and supervisors.

In~\cite{Lin2018}, 
 a formal formulation of the supervisor (augmented with a monitoring mechanism), the actuator enablement attacker, the attacked closed-loop system and resilient supervisor has been provided. The attacker is assumed to observe no more than the supervisor does on the execution of the plant. However, the attacker can eavesdrop the control commands issued by the supervisor. In this common setting, the control commands issued by the supervisor could be used to refine the knowledge of the attacker about the execution of the plant. The attacker can
modify each control command on a specified subset of attackable events. The attack
principle of the actuator attacker is to remain covert until it can establish a successful
attack and lead the attacked closed-loop system into generating certain damaging
strings. In~\cite{Lin2018}, a notion of attackability has been identified as a characterizing condition for the existence of a successful actuator enablement attacker, under a normality assumption imposed on the actuator attackers and supervisors; an algorithm for the verification of attackability has also been provided~\cite{Lin2018}. However, in~\cite{Lin2018}, the problem of synthesis of resilient supervisors against actuator enablement attacks has not been addressed.
 
Instead of directly synthesizing a supervisor that is resilient against actuator enablement attacks, it is possible to obfuscate an insecure supervisor to make it resilient, while preserving the behavior of the original closed-loop system. This is motivated by the following simple observation. The attacker can observe information from both the execution of the closed-loop system and the control commands issued by the supervisor. For any behavior-preserving supervisor, it cannot influence\footnote{For any behavior-preserving supervisor, by definition the behavior of the closed-loop system stays the same.} what the actuator attacker will observe from the execution of the closed-loop system but can partially determine what information the attacker can obtain\footnote{The supervisor is required to be behavior-preserving and thus  cannot issue arbitrary control commands.} from the control commands it issued. In order to be secure, a supervisor should leak as few information as possible in its issued control commands to any potential attacker. %This is practically feasible since two different control commands issued by the supervisor could have the same effect on the next step execution of the plant.

 %A basic 
 %An additional feature of the synthesized resilient supervisor, if it exists, is that it has exactly the minimum number of states among the set of all the resilient and behavior-preserving supervisors. 
 The main contributions of this paper are as follows. 
\begin{enumerate}
    \item We propose the problem of synthesis of resilient and behavior-preserving supervisors, i.e., the supervisor obfuscation problem, in the setting where there is a control command eavesdropping actuator enablement attacker. The supervisor that is provided as input to our problem may have been synthesized conforming to different kinds of complicated specifications and by using advanced synthesis techniques, but is insecure against actuator attack. Our approach avoids starting from scratch and developing a new resilient supervisor synthesis algorithm for each type of specification. An application of this approach is to first synthesize a supervisor that takes care of all the safety specifications, except for the resilience  property; the  insecure supervisor can then be  obfuscated to become resilient, if it exists.
    \item The problem of computing behavior-preserving supervisor is formulated as the problem of computing separating finite state automata~\cite{D12}, but with additional constraints such as controllability and observability, which can be efficiently reduced to the boolean satisfiability problem (SAT) and efficiently solved using modern SAT solvers.
    \item The algorithmic solution for the problem of computing behavior-preserving supervisors is used as a subroutine for the supervisor obfuscation problem. The obfuscated supervisor computed using our algorithm, if it exists, has exactly the minimum number of states among the set of all the resilient and behavior-preserving supervisors.
\end{enumerate}
The problem of computing behavior-preserving supervisors is essentially a flexible version of the supervisor reduction problem~\cite{VW86, SW04, S16}, the only differences being that there is no requirement on the state size of the computed supervisor and the control constraint of the computed supervisor does not have to be the same as that of the original supervisor. In this sense, the problem of computing behavior-preserving supervisors is a simple generalization of the supervisor reduction problem\footnote{However, the plants and supervisors used in this work are non-marking. The technique developed in this work can be easily extended to the marking case, which can be used for solving the standard supervisor reduction problem.}. The current implementation of the supervisor reduction procedure is not flexible and well tuned enough to be used as a solution to our problem, compared with the SAT-based approach. Apart from the SAT-solving technique, automaton learning and SMT solving~\cite{D12},~\cite{CFCTW09} can also be adopted to solve the problem. This work is mainly inspired by the work of~\cite{D12} on computing minimal separating finite state automata and the reduction technique developed in~\cite{D12} is adapted for solving the supervisor obfuscation problem.%, whose techniques are adopted for solving the supervisor obfuscation problem.

The paper is organized as follows. Section II is devoted to the general preliminaries; in addition, we recall the attack architecture, system formulation and definition of attackability provided  in~\cite{Lin2018}. Then, in Section III, we propose the problem of synthesis of resilient and behavior-preserving supervisors, i.e., the supervisor obfuscation problem, and then provide an algorithm to solve it. Finally, in Section IV, we conclude the paper and discuss some future research directions.

\section{Preliminaries}
In this section, we shall provide the basic preliminaries that are necessary to understand our paper. 
We assume the reader to be familiar with the basic theories of formal languages, finite automata and supervisory control~\cite{WMW10},~\cite{HU79}. Some additional notations and terminologies are introduced in the following. After that, we recall the attack architecture, system formulation and the definition of attackability in~\cite{Lin2018} that are specific to our setting.

Let $A$ and $B$ be any two sets. We write $A\backslash B$ to denote their set-theoretic difference, i.e., $A\backslash B:=\{x \in A \mid x \notin B\}$. Let $A \times B$ denote their Cartesian product. We write $|A|$ to denote the cardinality of $A$. A partial finite state automaton $G$ over alphabet  $\Sigma$ is a 5-tuple $(Q, \Sigma, \delta, q_0, Q_m)$, where $Q$ is the finite set of states, $\delta: Q \times \Sigma \longrightarrow Q$ the partial transition function\footnote{As usual, $\delta$ can also be viewed as a relation $\delta \subseteq Q \times \Sigma \times Q$.}, $q_0 \in Q$ the initial state and $Q_m$ the set of marked states.  $G$ is said to be a complete finite state automaton if $\delta: Q \times \Sigma \longrightarrow Q$ is a total function.  Let $L(G)$ (respectively, $L_m(G)$) denote the closed-behavior and the marked-behavior of $G$, respectively. When $Q_m=Q$, we also write $G=(Q, \Sigma, \delta, q_0)$ for simplicity, in which case we have $L_m(G)=L(G)$. In this work, whenever we talk about a plant, we mean a partial finite state automaton $G=(Q, \Sigma, \delta, q_0)$.  % and the marked behavior is denoted by $L_m(G)$. 
$G$ is said to be $n$-bounded if $|Q| \leq n$. For two finite state automata $G_1=(X_1, \Sigma_1, \delta_1, x_{1,0}), G_2=(X_2, \Sigma_2, \delta_2, x_{2,0})$, we write $G:=G_1 \lVert G_2$ to denote their synchronous product. Then, 
$G = (X := X_1\times X_2,\Sigma:=\Sigma_1\cup\Sigma_2,\delta = {\delta}_1 \times {\delta}_2,x_0:=(x_{1,0},x_{2,0}))$
where the (partial) transition map $\delta$ is defined as follows.
$ (\forall x = (x_1,x_2)\in X)(\forall \sigma \in \Sigma)\delta(x,\sigma)$ $$ :=\left\{
\begin{array}{rcl}
({\delta}_1(x_1,\sigma),x_2), && \text{if } {\sigma \in {\Sigma}_1}\backslash {\Sigma}_2 \\
(x_1,{\delta}_2(x_2,\sigma)), && \text{if } {\sigma \in {\Sigma}_2}\backslash {\Sigma}_1 \\
({\delta}_1(x_1,\sigma),{\delta}_2(x_2,\sigma)), && \text{if } {\sigma \in {\Sigma}_1}\cap {\Sigma}_2 \\
\end{array} \right. $$

%In the propositional logic , a formula $\phi$ consists of  and logical  connectives . If the variables occurring in $\phi$ are of special interest, we also write $\phi(x_1,\ldots,x_n)$. A model of $\phi$ is a mapping $\mathcal{M}:Var(\phi)\rightarrow \{0,1\}$ (0 representing false, 1 representing true) such that $\phi$ evaluates to be true if all variables $x_i$ in $\phi$ are substituted by $\mathcal{M}(x_i)$, written as $\mathcal{M} \models \phi$. {\color{red} this part can be improved to be more precise, but not necessarily needed}

Propositional formulas $\phi$ ~\cite{D12} ~\cite{what12} are constructed from (Boolean) variables by using logical connectives ($\wedge,\vee, \rightarrow,\neg$). The truth value of a propositional formula is determined by the variables' truth value.  A literal is a variable or its negation. A clause is a disjunction $l_1 \vee \ldots \vee l_n$ of literals. A formula in conjunctive normal form (CNF) is conjunction of clauses. Each propositional formula can be converted into an equivalent formula that is in conjunctive normal form. Let $Var(\phi)$ denote the set of variables (Boolean variables $x_0,\ldots,x_n$) occurring in $\phi$;  A model of $\phi$ is a mapping ${\bf M}$: $Var(\phi) \rightarrow \{ 0,1\}$ (0 representing False, 1 representing True) such that $\phi$ is evaluated to be True if all variables $x_i$ in $\phi$ are substituted by $\text{M}(x_i)$.

%{\color{red} def. of a satisfying assigment needs improvement}

%That is, Assuming $\phi$ is a mapping function $\mathcal{M}:Var(\phi)\rightarrow \{0,1\}$ (0 representing false, 1 representing true) such that $\phi$ evaluates to be true if all variables $x_i$ in $\phi$ are true. % {\color{red} conjunctive normal form: variables, literals, clauses, formulas..}

%{\color{red}
%partial model, model of sat formula
%we write $G_1=G_2$ if $L_m(G_1)=L_m(G_2)$ and $L(G_1)=L(G_2)$
%synchronous product. as refinement of state space of transition structure.
%automaton completion
%}

A control constraint $\mathcal{A}$ over $\Sigma$ is a tuple ($\Sigma_c,\Sigma_o$), where  $\Sigma_c\subseteq\Sigma$ denotes the subset of controllable events and $\Sigma_o\subseteq\Sigma$ denotes the subset of  observable events.
Let $\Sigma_{uo}:=\Sigma\backslash\Sigma_o$ denote the subset of unobservable events and $\Sigma_{uc}:=\Sigma\backslash\Sigma_c$ the subset of uncontrollable events. For each sub-alphabet $\Sigma' \subseteq \Sigma$, the natural projection $P: \Sigma^* \rightarrow \Sigma'^*$ is defined and naturally extended to a mapping between languages~\cite{WMW10}. %A supervisor on $G$ w.r.t. control constraint $(\Sigma_c, \Sigma_o)$ is a map $V: P_{o}(L(G)) \rightarrow \Gamma$, where $P_o: \Sigma^* \rightarrow \Sigma_o^*$ is the natural projection and $\Gamma:=\{\gamma \subseteq \Sigma \mid \Sigma_{uc} \subseteq \gamma\}$. $V(w)$ is the control command issued by the supervisor after observing $w \in P_{o}(L(G))$. $V/G$ denotes the closed-loop system of $G$ under the supervision of $V$. The closed-behavior $L(V/G)$ of $V/G$ is inductively defined as follows:
%\begin{enumerate}
%\item $\epsilon \in L(V/G)$,
%\item if $s \in L(V/G)$,  $\sigma \in V(P_{o}(s))$ and $s\sigma \in L(G)$, then $s\sigma \in L(V/G)$,
%\item no other strings belong to $L(V/G)$.
%\end{enumerate}

%For any two strings $s, t \in \Sigma^*$, we write $s \leq t$ (respectively, $s<t$) to denote that $s$ is a prefix (respectively, strict prefix) of $t$. For any language $L \subseteq \Sigma^*$, $\overline{L}$ is used to denote the prefix-closure of $L$. 

{\bf Attack Architecture}: We consider the architecture for actuator attack shown in Fig.~\ref{fig:ring}. The plant $G$ is under the control of a partially observing supervisor $V$ w.r.t. $(\Sigma_c, \Sigma_o)$.  In addition, there exists an actuator attacker $A$ that can observe the execution of the plant and eavesdrop the control commands issued by the supervisor. We assume
that each time when the supervisor observes an event (and makes a state transition), it issues a new control command that can be intercepted by the attacker. It can modify the control command $\gamma$ issued by the supervisor on a designated subset of attackable events $\Sigma_{c, A} \subseteq \Sigma_c$, each time when it intercepts a control command. The plant $G$ follows the modified control command $\gamma'$ instead of $\gamma$. The attacker can observe events in $\Sigma_{o, A} \subseteq \Sigma_o$ in the execution of the plant.  $(\Sigma_{c, A}, \Sigma_{o, A})$ is said to be an attack constraint over $\Sigma$. We assume $\Sigma_{c} \subseteq \Sigma_o$ and $\Sigma_{c, A} \subseteq \Sigma_{o, A}$. That is, both the actuator attackers and supervisors are assumed to be normal~\cite{Lin2018}.
The supervisor is augmented with a monitoring mechanism that monitors the execution of the closed-loop system (under attack). The supervisor has a mechanism for halting the execution of the closed-loop system after discovering an actuator attack. The goal of the attacker is to drive the attacked closed-loop system into executing specific damaging strings outside the controlled behavior, without risking itself being discovered by the supervisor before it causes damages.    

%As we shall see soon, since the attacker aims to drive the close-loop system into executing certain damaging strings, it is never in the interest of the attacker to cause system deadlock (prematurely). {\color{red} check} {\color{red} ; however, the attacker can neither delete existing control commands nor insert new control commands\footnote{Intuitively, vulnerable actuators give rise to attackable events. The attacker has no means of attacking actuators that are not vulnerable.}}

\begin{figure}[h]
\centering
\hspace*{-1mm}
\includegraphics[width=2.6in, height=1.4in]{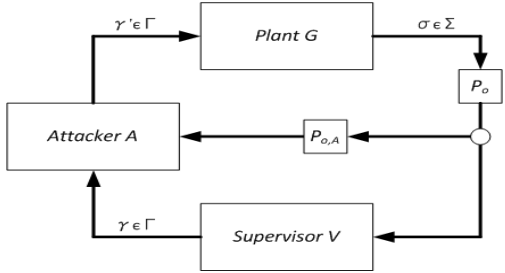} 
\caption{The architecture for actuator attack}% with $\Sigma_{i_1} \cup \Sigma_{i_2}$
\label{fig:ring}
\end{figure}

{\bf System Formulation:}
\subsubsection{Supervisor}
A supervisor\footnote{The standard definition of a supervisor $V: P_o(L(G)) \rightarrow \Gamma$ requires that
the supervisor shall also apply control for those strings $w \in P_o(L(G))-P_o(L(V/G))$
that cannot be observed in the normal execution of the closed-loop system. From the
point of view of the supervisor, it will conclude the existence of an (actuator) attacker
and then halt the execution of the closed-loop system the first time when it observes
some string $w \notin P_o(L(V/G))$. Under this interpretation, the supervisor is augmented
with a monitoring mechanism for the detection of actuator attack and does not control
outside $P_o(L(V/G))$. The renewed definition of $V$ is not circular.   See~\cite{Lin2018} for more detailed explanation.} $V$ on $G$ w.r.t. $(\Sigma_c, \Sigma_o)$ is effectively a map $V: P_{o}(L(V/G)) \rightarrow \Gamma$, where $\Gamma=\{\gamma \subseteq \Sigma \mid \Sigma_{uc} \subseteq \gamma\}$ and $L(V/G)$ is the closed behavior of the closed-loop system $V/G$~\cite{WMW10}. Once any string $w \notin P_o(L(V/G))$ is observed by the supervisor, the closed-loop system is immediately halted, since an attack has been discovered by the supervisor. For any $w \in P_o(L(V/G))$, we have $V(w) \supseteq \Sigma_{uc} \supseteq \Sigma_{uo}$ by the definition of $\Gamma$ and since  $\Sigma_c \subseteq \Sigma_o$. A supervisor $V$ (on $G$) w.r.t. control constraint $(\Sigma_c,\Sigma_o)$ is realized by a partial finite state automaton $S = (X,\Sigma,\xi,x_0)$ that satisfies the controllability and observability constraints ~\cite{B1993}: 
\begin{enumerate}
    \item [{\bf C})] (controllability) for any state $x\in X$ and any uncontrollable event $\sigma \in \Sigma_{uc}$, $\xi(x,\sigma)!$,
    \item [{\bf O})] (observability) for any state $x\in X$ and any unobservable event $\sigma \in \Sigma_{uo}$, $\xi(x,\sigma)!$ implies $\xi(x,\sigma)=x$,
\end{enumerate}
where we have $L(V/G) = L(S \parallel G)$ and, for any $s \in L(G)$ such that $P_o(s) \in P_o(L(V/G))$, 
$V(P_o(s)) = \{ \sigma \in \Sigma \mid \xi(\xi(x_0,s_0),\sigma)!,s_0\in L(S)\cap P_o^{-1}P_o(s) \}$. For a normal supervisor $S$ w.r.t. control constraint $(\Sigma_c,\Sigma_o)$, the observability constraint is then reduced to: for any state $x \in X$ and any unobservable event $\sigma \in \Sigma_{uo}$, $\xi (x,\sigma) = x$.
\subsubsection{Damage-inflicting set}
\label{subsubs: Damage}
The goal of the actuator attacker is to drive the attacked closed-loop system into executing certain damaging strings. Let $L_{dmg} \subseteq L(G)$ be some regular language over $\Sigma$ that denotes a so-called ``damage-inflicting set", where each string is a damaging string. We  require that\footnote{This is not restrictive at all, as we can easily constrain $V$ in its synthesis such that $L(V/G) \subseteq L_{dmg}^c$.} $L_{dmg} \cap L(V/G)=\varnothing$, that is, no string in $L(V/G)$ could be damaging. Then, we have that $L_{dmg} \subseteq L(G)-L(V/G)$. We remark that $L_{dmg}$ is a general device for specifying what are damaging strings and may not correspond to the specification that is used for synthesizing $V$. $L_{dmg}$ is given by the damage automaton $H=(Z, \Sigma, \eta, z_0, Z_m)$, i.e., $L_{dmg}=L_m(H)$. We require $H$ to be a complete automaton and thus $L(H)=\Sigma^*$. 

\subsubsection{Actuator Attacker}
\label{subsubs: AA}
The attacker's observation sequence is simply a string in $((\Sigma_{o, A} \cup \{\epsilon\}) \times \Gamma)^*$. The attacker's observation sequence solely depends
on the supervisor’s observation sequence $P_o(s) \in P_o(L(V/G))$, including the observation on the execution of the plant and the control commands issued by the supervisor.  $\hat{P}_{o, A}^{V}: P_o(L(V/G)) \rightarrow ((\Sigma_{o, A} \cup \{\epsilon\}) \times \Gamma)^*$ is a function that maps supervisor's observation sequences to attacker's observation sequences, where, for any $w=\sigma_1\sigma_2\ldots \sigma_n \in P_o(L(V/G))$, $\hat{P}_{o, A}^{V}(w)$ is defined to be
\begin{center}
 $(P_{o, A}(\sigma_1), V(\sigma_1))(P_{o, A}(\sigma_2), V(\sigma_1\sigma_2))\ldots$ \\ $(P_{o, A}(\sigma_n), V(\sigma_1\sigma_2 \ldots \sigma_n)) \in ((\Sigma_{o, A} \cup \{\epsilon\}) \times \Gamma)^*$.
\end{center}
An attacker on $(G, V)$ w.r.t. attack constraint $(\Sigma_{c, A}, \Sigma_{o, A})$ is a function $A: \hat{P}_{o, A}^{V}(P_o(L(V/G))) \rightarrow \Delta \Gamma$, where $\Delta \Gamma=2^{\Sigma_{c, A}}$. Here, $\Delta \Gamma$ denotes the set of all the possible attack decisions that can be made by the attacker on the set $\Sigma_{c, A}$ of attackable events. Intuitively, each $\Delta \gamma \in \Delta \Gamma$ denotes the set of enabled attackable events that are determined by the attacker. For any $P_o(s) \in P_o(L(V/G))$, the attacker determines the set $\Delta \gamma=A(\hat{P}_{o, A}^{V}(P_o(s)))$ of attackable events to be enabled based on its observation $\hat{P}_{o, A}^{V}(P_o(s))$. 

Now, given the supervisor $V$ and the actuator attacker $A$, we could lump them together into an equivalent (attacked) supervisor $V_A: P_o(L(V/G)) \rightarrow \Gamma$ such that, for any string $w \in P_o(L(V/G))$ observed by the attacked supervisor, the control command issued by $V_A$ is $V_A(w)=(V(w)-\Sigma_{c, A}) \cup A(\hat{P}_{o, A}^{V}(w))$. The plant $G$ follows the control command issued by the attacked supervisor $V_A$ and the attacked closed-loop system is denoted by $V_A/G$. The definition of the (attacked) closed-behavior $L(V_A/G)$ of $V_A/G$ is inductively defined as follows.

\begin{enumerate}
\item $\epsilon \in L(V_A/G)$,
\item if $s \in L(V_A/G)$, $P_o(s) \in P_o(L(V/G))$, $\sigma \in V_A(P_{o}(s))$ and $s\sigma \in L(G)$,  then $s\sigma \in L(V_A/G)$,
\item no other strings belong to $L(V_A/G)$.
\end{enumerate}
The goal of the actuator attacker is formulated as follows:
$L(V_A/G) \cap L_{dmg}\neq \varnothing$, i.e., at least some damaging string can be generated before the execution of the closed-loop system is halted. If the goal is achieved, then we say $A$ is a successful actuator attacker on $(G, V)$ w.r.t. $(\Sigma_{c, A}, \Sigma_{o, A})$ and $L_{dmg}$. Under the normality assumption, without loss of generality, we can assume $A$ is an enablement attacker~\cite{Lin2018}, i.e., $A(\hat{P}_{o, A}^{V}(w)) \supseteq V(w) \cap \Sigma_{c, A}$, for any $w \in P_o(L(V/G))$.

{\bf Attackability:} We now recall the definition of attackability, which plays an important role in characterizing the existence of a successful actuator attacker.
\begin{definition}
$(G, V)$ is said to be attackable w.r.t. $(\Sigma_{c, A}, \Sigma_{o, A})$ and $L_{dmg}$ if there exists a string $s \in L(V/G)$ and an attackable event $\sigma \in \Sigma_{c, A}$, such that 
\begin{enumerate}
\item $s\sigma \in L_{dmg}$
\item for any $s' \in L(V/G)$, $\hat{P}_{o, A}^{V}(P_o(s))=\hat{P}_{o, A}^{V}(P_o(s'))$ and $s'\sigma \in L(G)$ together implies $s'\sigma \in L_{dmg}$. 
\end{enumerate}
\end{definition}
We recall the following result~\cite{Lin2018}.  
\begin{theorem}
There exists a successful actuator attacker on $(G, V)$ w.r.t. $(\Sigma_{c, A}, \Sigma_{o, A})$ and $L_{dmg}$ iff $(G, V)$ is attackable w.r.t $(\Sigma_{c, A}, \Sigma_{o, A})$ and $L_{dmg}$.
\end{theorem}

An algorithm for (non-)attackability verification is available in~\cite{Lin2018}.

\section{ Supervisor Obfuscation Problem}
In this section, we shall propose and address the problem of supervisor obfuscation, which is an approach for the synthesis of resilient supervisors against actuator attackers, while preserving desired behavior of the original closed-loop system.
\subsection{Problem Formulation}
A supervisor $V'$ is said to be an {\em obfuscation} of $V$ (against actuator attacks) on $G$ w.r.t. $(\Sigma_{c, A}, \Sigma_{o, A})$ and $L_{dmg}$ if the following two conditions are satisfied: 1) there is no successful actuator attacker on $(G, V')$ w.r.t. $(\Sigma_{c, A}, \Sigma_{o, A})$ and $L_{dmg}$, and 2) $L(V/G)=L(V'/G)$. In general, we do not require $V'$ to be over the same control constraint as $V$ and, in practice, $V$ is insecure (for us to be interested in the supervisor obfuscation problem). It is worth noting that $L(V'/G)=L(V/G)$ does not imply the equi-attackability of $(G,V)$ and $(G,V')$ because in general $V\neq V'$ and then ${\hat P}_{o,A}^V\neq {\hat P}_{o,A}^{V'}$ (c.f. Definition 1). This shall be  illustrated by the next example.
\begin{example}
We shall provide a simple  supervisor obfuscation example below.
Let us consider the plant $G$ and the supervisor $S$ shown in Fig. 2. The colored state 8 is the only bad state to avoid. $\Sigma = \{a,b,c,d,a'\}, \Sigma_o = \{a,c,d, a'\}, \Sigma_c = \Sigma_o , \Sigma_{c,A} = \{a'\}, \Sigma_{o,A} = \{c, a'\}$. We can check that $S$ has ensured safety,  without the presence of an attacker. The damage automaton $H$ (with the dump state and the corresponding transitions being omitted) is also shown in Fig. 2, where state 8 is the only marked state. However, $(G, S)$ can be attacked w.r.t. $(\Sigma_{c, A}, \Sigma_{o, A})$ and $H$. When the plant generates string $\omega =acd$, the supervisor knows the current state is exactly state 6. However, event $d$ is unobservable to the attacker; before the execution of event $d$. the attacker will firstly estimate the current state should be state 5 or state 6. Then, with the eavesdropped control command information from the supervisor after $d$ is executed, it can know that event $d$ has occurred because $V(ac)=\{a, b, d\}$ in supervisor state 3 while $V(acd) = \{b\}$ in supervisor state 4. Then, it knows the plant is in state 6 and thus an attack on event $a'$ can be established. 

We have also shown another supervisor $S'$ in Figure 2, which also has ensured the safety of the closed-loop system. We can easily check that $L(S \lVert G)=L(S' \lVert G)$ and thus $S'$ is behavior-preserving. To prevent the attacker from detecting the occurrence of event $d$, compared with $S$, the supervisor $S'$ self-loops event $d$ at state 3 instead and thus $V'(ac)=V'(acd)= \{a, b, d\}$. Effectively, the attacker cannot determine whether it is event $d$ or $a$ that occurs, even if it eavesdrops a control command and knows that a transition has occurred. Therefore, the attacker will not take the risk to carry out the attack and $(G,S')$ is not attackable w.r.t. $(\Sigma_{c, A}, \Sigma_{o, A})$ and $H$. 
%Once the faked self-loop event transition added is to state 3 of obfuscated supervisor $S'$, attacker cannot distinguish state 5 and 6 in plant $G$ anymore because $V'(ac) = {a,d}$ in both of them. Therefore, it won't take the risk to conduct the attack until it is one hundred percent successful. (G,S') is not attackable w.r.t. $(\Sigma_{c, A}, \Sigma_{o, A})$ and $H$. 
\end{example}
\begin{figure}[h]
\centering
\hspace*{-1mm}
\includegraphics[width=3in, height= 3.2 in]{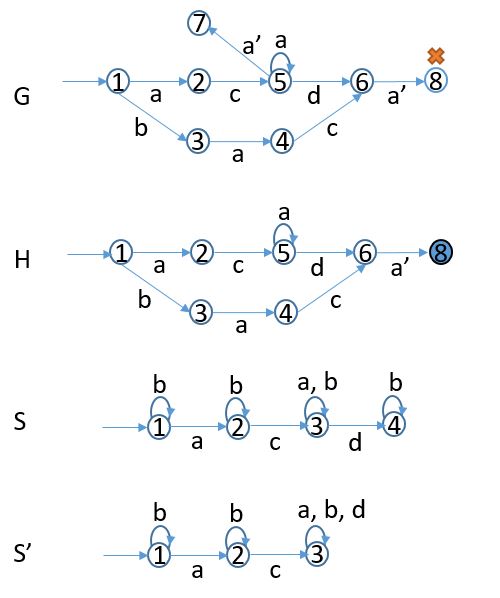} 
\caption{The plant $G$, the damage automaton $H$, the original supervisor $S$ and  the obfuscated supervisor $S'$}% with $\Sigma_{i_1} \cup \Sigma_{i_2}$
\label{fig:ring}
\end{figure}

In the rest, we implicitly fix the alphabet $\Sigma$ and use automata representations as problem inputs. We are now ready to state the main problems.

\begin{problem}
Given a plant $G$, a supervisor $S$, a control constraint $\mathcal{A}=(\Sigma_{c}, \Sigma_{o})$, an attack constraint $(\Sigma_{c, A}, \Sigma_{o, A})$, a damage automaton $H$ and a positive integer $n$, does there exist an $n$-bounded supervisor $S'$  over $\mathcal{A}$ such that 1) $L(S'\parallel G)=L(S\parallel G)$ and 2) $(G, S')$ is not attackable w.r.t. $(\Sigma_{c, A}, \Sigma_{o, A})$ and $H$.
\end{problem}
Problem 1 asks whether there exists a resilient and behavior-preserving supervisor over $\mathcal{A}$ of state size no more than $n$. The optimization version of Problem 1, which is given as Problem 2, is the supervisor obfuscation problem that we intend to solve. 

\begin{problem}
Given a plant $G$, a supervisor $S$, a control constraint $\mathcal{A}=(\Sigma_{c}, \Sigma_{o})$, an attack constraint $(\Sigma_{c, A}, \Sigma_{o, A})$, a damage automaton $H$, compute a supervisor $S'$  over $\mathcal{A}$, if it exists, such that 1) $L(S'\parallel G)=L(S\parallel G)$, 2) $(G, S')$ is not attackable w.r.t. $(\Sigma_{c, A}, \Sigma_{o, A})$ and $H$, and 3) $S'$ has the minimum number of states among the set of all the supervisors over $\mathcal{A}$ that satisfy both 1) and 2).
\end{problem}
We shall develop an algorithmic solution in Section~\ref{sec: ASO} that can be used for solving both Problem 1 and Problem 2. We have the following remark.
\begin{remark}
Since $S$ and $G$ are given, we can compute $K=L(S \lVert G)$. Thus, we do not need to know the control constraint of the original supervisor $S$ and the computed supervisor $S'$ can be over a different control constraint than that of $S$.
\end{remark}

%From the supervisor's point of view, even if some issued control commands are known by the attacker, it is still necessary to hide its behaviour as much as possible to confuse the attacker  therefore to avoid further damage. This section will illustrate how to find a equivalent automaton which has the same language and closed-loop behaviour as the original one.

\subsection{Algorithm for Supervisor Obfuscation}
\label{sec: ASO}
Our proposed solution for solving the supervisor obfuscation problem is given in Algorithm 1. In particular, 
 Algorithm \textbf{SUPBP}($G,S,\mathcal{A},n$) is used to generate the stack $I$ of all the supervisors that are behavior-preserving with respect to $S$ and have state size $n$. This procedure can be completed efficiently with the help of an ALL-SAT solver~\cite{GSY04},~\cite{TS15}. Algorithm \textbf{NA}($G, S, H$) is used to verify the  non-attackability of $(G, S')$ w.r.t. $(\Sigma_{c, A}, \Sigma_{o, A})$ and $H$. The details of Algorithm \textbf{SUPBP}($G,S,\mathcal{A},n$) and Algorithm \textbf{NA}($G, S', H$) will be explained in the following subsections. 
 
 The algorithm for supervisor obfuscation starts by searching for behavior-preserving supervisors of state size $n=1$. If there is no such behavior-preserving supervisor, then we increment state size $n$ and redo the search. If there is a behavior-preserving supervisor of state size $n$, then we store the set of all behavior-preserving supervisors of state size $n$ in stack $I$, obtained by using Algorithm \textbf{SUPBP}($G, S, \mathcal{A}, n$). For each behavior-preserving supervisor in the stack, we check the non-attackability of the closed-loop system using Algorithm \textbf{NA}($G, S, H$). If there exists a behavior-preserving supervisor in the stack so that the closed-loop system is non-attackable, we have found a resilient and behavior-preserving supervisor, which is guaranteed to be of the minimum number of states. If the closed-loop system associated with each behavior-preserving supervisor in the stack is attackable, then we increment $n$ and redo the search. To prevent the algorithm from searching infinitely (by increasing $n$ indefinitely), we can impose a bound on the largest $n$ that we would like to search. Typically, we choose the largest $n$ to be $|X|$ (that is, we would like to synthesize a resilient and behavior-preserving supervisor $S'$ of smaller state size than that of $S$) or some small multiples of $|X|$. 
We remark that the bisection method can be used to improve the performance of the searching process.

\begin{algorithm}[h]
\caption{Algorithm for supervisor obfuscation}
\hspace*{\algorithmicindent} \textbf{Input} Plant $G=(Q,\Sigma,\delta,q_0)$, supervisor $S=(X,\Sigma,\xi,x_0)$, control constraint $\mathcal{A}$, attack constraint ($\Sigma_{c,A}$, $\Sigma_{o,A}$), damage automaton $H= (Z,\Sigma,\eta,z_0, Z_m)$. \\
\hspace*{\algorithmicindent} \textbf{Output} Supervisor $S'=(X',\Sigma,\xi',x_0')$ %or NO$\_$SOLUTION
\begin{algorithmic}[1]
\State $n:=1$
%\State \textbf{If}  $n> N$, \textbf{return} NO$\_$SOLUTION
\State Compute the stack of behavior-preserving  supervisors $I = \textbf{SUPBP} (G,S,\mathcal{A},n)$ of state size $n$
\State \textbf{If}  $I= \varnothing$ 
\State \quad \quad $n:=n+1$, \textbf{goto} step 2
\State \textbf{While} $I \neq \varnothing$ 
\State \quad Let $S'=I.pop()$ 
\State \quad Compute \textbf{NA}$(G,S',H)$
\State \quad \textbf{If} \textbf{NA}$(G,S',H)$ = \text{TRUE}
\State \quad \quad \textbf {return} $S'$
\State $n:=n+1$, \textbf{goto} step 2 
\end{algorithmic}
\end{algorithm}

 The formula $\phi_n^{\overline{G\downarrow S},\mathcal{A}}$ used in Algorithm \textbf{SUPBP}($G,S,\mathcal{A},n$) states the existence of an $n$-bounded supervisor over $\mathcal{A}$ that is behavior-preserving w.r.t. $S$ over $G$. A satisfying assignment ${\bf M}$ of $\phi_n^{\overline{G\downarrow S},\mathcal{A}}$ can be easily converted to an $n$-bounded behavior-preserving supervisor $S'^{{\bf M}}$ on $G$ over $\mathcal{A}$. Only when the set of reachable states of $S'^{{\bf M}}$ is of cardinality $n$, we can then add $S'^{{\bf M}}$ to the stack $I$, in Algorithm \textbf{SUPBP}($G,S,\mathcal{A},n$), to ensure that $I$ only has behavior-preserving supervisors of state size exactly $n$. This avoids repetition of non-attackability verification carried out in Step 7 of Algorithm 1, as supervisors found by solving $\phi_n^{\overline{G\downarrow S},\mathcal{A}}$ are only guaranteed to be $n$-bounded.

\begin{algorithm}
\caption{Algorithm \textbf{SUPBP}($G,S,\mathcal{A},n$)}
\hspace*{\algorithmicindent} \textbf{Input:} Plant $G=(Q,\Sigma,\delta,q_0)$, control constraint $\mathcal{A}$, supervisor $S = (X,\Sigma,\xi,x_0)$ and a positive integer $n$ \\
\hspace*{\algorithmicindent} \textbf{Output:}  Stack $I$ of behavior-preserving supervisors of state size $n$ over $\mathcal{A}$. 
\begin{algorithmic}[1]
\State Construct the boolean formula $\phi_n^{\overline{G\downarrow S},\mathcal{A}}$
\State Let $I=\varnothing$
\State \textbf{If} \quad {$\phi_n^{\overline{G \downarrow S},\mathcal{A}}$ is satisfiable}
\State \quad \textbf{for} each satisfying assignment ${\bf M}$ of $\phi_n^{\overline{G\downarrow S},\mathcal{A}}$
\State \quad \quad Construct the supervisor $S'^{{\bf M}}$ that corresponds to ${\bf M}$ 
\State \quad \quad If $S'^{{\bf M}}$ is of state size $n$ 
\State \quad \quad \quad Let $I:=I.push(S'^{{\bf M}})$
\State \textbf{return} $I$
\end{algorithmic}
\end{algorithm}

\begin{algorithm}
\caption{Algorithm \textbf{NA}($G,S',H$)}
\hspace*{\algorithmicindent} \textbf{Input:} 
Plant $G=(Q,\Sigma,\delta,q_0)$, control constraint $\mathcal{A}$, supervisor ${S'}=(X',\Sigma,\xi',x_o') $, damage automaton $H = (Z,\Sigma,\eta,z_0, Z_m)$ \\
\hspace*{\algorithmicindent} \textbf{Output:} TRUE or FALSE
\begin{algorithmic}[1]
\State Compute the annotated supervisor ${S'}^A$ of $S'$ 
\State Compute the general synchronous product $GP(G,{S'}^A,H)$ of $G$, $S'^A$ and $H$
\State Compute the automaton $SUB(GP(G,S'^A,H))$ with a labeling function $Lf$
\State \textbf{If} for  every  reachable  state $y$ in $SUB(GP(G,S'^A,H))$ it holds that $Lf(y)=\varnothing$
\State \quad \textbf{return} TRUE
\State \textbf{else} 
\State \quad \textbf{return} FALSE
\end{algorithmic}
\end{algorithm}

%{\color{red} Instead of supervisor reduction, we shall say behavior-preserving supervisors, as the state size may be larger than the original supervisor.}

%It is known that the problem of computing minimal separating finite state automaton can be reduced to SAT solving. It is not surprising that the problem of supervisor reduction,  and $n\geq 1$
%partial finite state automaton $G$ over   and finite state supervisor $M$ over $\mathcal{A}$ over $\Sigma$ , where $S=G \lVert M$

{\bf Computing Behavior-Preserving Supervisors:}
We now show that the problem of computing behavior-preserving supervisors, which is given below, is equivalent to the problem of computing separating finite state automata under controllability and observability constraints. This shall allow us to borrow the technique developed for computing minimal separating finite state automata to solve our problem. Essentially, for any given plant $G$, supervisor $S$ and control constraint $\mathcal{A}$, we shall construct boolean formulas $\phi_n^{\overline{G\downarrow S},\mathcal{A}}$ that state the existence of an $n$-bounded supervisor over $\mathcal{A}$ that is behavior-preserving w.r.t. $S$ over $G$.

%{\color{red} The rest needs to be changed. For our plant, we use partial finite state automaton, need to first add a dump state to make it complete. Add markings to all states that are not dump state; then can use exactly the definition of separating finite state automaton in Neider's paper.}

%{\color{red} need to define $\equiv$, i.e., behavior preserving}

\begin{problem}[Existence of $n$-Bounded  Behavior-Preserving Supervisors]
Given a plant $G$, a control constraint $\mathcal{A}=(\Sigma_c, \Sigma_o)$ and a supervisor $S$, determine the existence of an $n$-bounded supervisor $S'$ over $\mathcal{A}$ such that $L(S \lVert G)=L(S' \lVert G)$.
\end{problem}

%First, the relationship between the problem of optimal supervisor reduction and the problem of computing a minimal separating partial finite automaton will be explained. The definitions of separating automaton and partial finite state automaton are needed.
We shall now introduce the well-known notion of a separating finite state automaton~\cite{D12}, which works in the context of complete finite state automata.
\begin{definition}[Separating Finite State Automaton]
Given any (complete) finite state automaton $C$ and any two languages $L_1, L_2$, if $L_1 \subseteq L_m(C)$ and $L_m(C) \cap L_2=\varnothing$, then $C$ is said to be a separating finite state automaton for the pair $(L_1, L_2)$.
\end{definition}
Intuitively, $C$ is a witness of the disjointness of $L_1$ and $L_2$. We make a simple observation in Lemma~\ref{lemma: obser}, which relates the notion of a  behavior-perserving supervisor with the notion of a separating finite state automaton. In order to relate these two notions, we need to convert the plant and the supervisor\footnote{We assume the non-degenerate and practically relevant case where both supervisors and plants are strictly partial finite state automata. The degenerate case can also be treated in a similar fashion.}, to a complete finite state automaton by adding a dump state, so that a partial finite state automaton $P=(W, \Sigma, \pi, w_0)$ becomes a complete finite state automaton $\overline{P}=(W \cup \{w_d\}, \Sigma, \overline{\pi}, w_0, W)$, where the subscript $d$ is always used with dump state and $\overline{\pi}=$
\begin{center}
$\pi \cup (\{w_d\} \times \Sigma \times \{w_d\}) \cup \{(w, \sigma, w_d) \mid \pi(w, \sigma)$ is undefined$, w \in W, \sigma \in \Sigma\}$
\end{center}
After the above completion process, we have $L(P)=L_m(\overline{P})$. It is also  straightforward to recover $P$  from $\overline{P}$, by removing the dump state (and the corresponding transitions). In the rest, we shall work on $\overline{G}$, $\overline{S}$ and $\overline{S'}$ instead. In particular,  $\overline{G}$ and $\overline{S}$ can be obtained from $G$ and $S$ by the above completion process; $S'$ is the behavior-preserving supervisor that we would like to compute and we only need to obtain $\overline{S'}$ first and then remove the unique dump state of $\overline{S'}$ to recover $S'$. Then, our goal is to compute $\overline{S'}$ in the following.

Based on the above discussion, we  need to ensure $L_m(\overline{G}) \cap L_m(\overline{S'})=L_m(\overline{G}) \cap L_m(\overline{S})$, which is equivalent to $L(G) \cap L(S')=L(G) \cap L(S)$. Now, we are ready to state Lemma~\ref{lemma: obser}.
%For transferring the non-marking automaton into the marked language, marked states $Q_m$ are supposed to be added. Therefore, We transferred the 4-tuple automaton ($\overline{Q},\Sigma,\overline{\delta},q_o,Q_m$) by adding one dump state and all the original states are regarded as marked state. 

\begin{lemma}
\label{lemma: obser}
Let $\overline{G}, \overline{S}$ and $\overline{S'}$ be any given complete finite state automata.  $L_m(\overline{G}) \cap L_m(\overline{S'})=L_m(\overline{G}) \cap L_m(\overline{S})$ if and only if 
\begin{enumerate}
\item $L_m(\overline{G}) \cap L_m(\overline{S}) \subseteq L_m(\overline{S'})$
\item $L_m(\overline{S'}) \cap (L_m(\overline{G}) \backslash L_m(\overline{S}))=\varnothing$
\end{enumerate}
\end{lemma}

According to the above result and discussion, to solve Problem 3 we shall compute an $(n+1)$-bounded $\overline{S'}$ (since $S'$ is $n$-bounded) that satisfies the following conditions: 
\begin{enumerate}
\item $\overline{S'}$ separates $(L_m(\overline{G}) \cap L_m(\overline{S}), L_m(\overline{G}) \backslash L_m(\overline{S}))$;
\item The $S'$ part of $\overline{S'}$ satisfies constraints {\bf (C)} and {\bf (O)}.
\end{enumerate}

We recall that $G=(Q, \Sigma, \delta,q_0)$ and $S=(X, \Sigma, \xi, x_0)$. 
It is straightforward to encode in a ``generalized" complete finite state automaton the two languages $L_m(\overline{G}) \cap L_m(\overline{S})$ and $L_m(\overline{G}) \backslash L_m(\overline{S})$, by using two different kinds of markings. Let $\overline{G \downarrow S}=(Y, \Sigma, \rho, y_0, Y_A, Y_B)$ denote such a generalized (complete) finite state automaton with two kinds of marking states $Y_A$ and $Y_B$. The set $Y_A \subseteq Y$ of marking states is used to recognize $L_m(\overline{G}) \cap L_m(\overline{S})$ and the set $Y_B \subseteq Y$ of marking states is used to recognize $L_m(\overline{G}) \backslash L_m(\overline{S})$. $\overline{G \downarrow S}$ can be obtained by computing the synchronous product of $\overline{G}$ and $\overline{S}$, with $Y_A=Q \times X$ and $Y_B=Q \times \{x_d\}$.

We adopt the technique of~\cite{D12} and provide a polynomial-time reduction from Problem 3 to the SAT problem; it then follows that a SAT solver could be deployed to solve Problem 3. In the high level, the idea of the reduction proceeds as follows: for any given instance of Problem 3 with input  $\overline{G}$, $\overline{S}$ and control constraint $\mathcal{A}$, we produce a propositional  formula $\phi_n^{\overline{G \downarrow S}, \mathcal{A}}$ such that $\phi_n^{\overline{G \downarrow S}, \mathcal{A}}$ is satisfiable if and only if there exists an $(n+1)$-bounded $\overline{S'}$  such that  $\overline{S'}$ separates $(L_m(\overline{G}) \cap L_m(\overline{S}), L_m(\overline{G}) \backslash L_m(\overline{S}))$. Moreover, each model of the formula $\phi_n^{\overline{G \downarrow S}, \mathcal{A}}$ can be directly translated to an $(n+1)$-bounded $\overline{S'}$ (correspondingly, an $n$-bounded $S'$) that solves the given instance. %Then, a SAT solver~\cite{what12} could be used to compute an optimally reduced supervisor, by increasing the value of $n$ until $\phi_n^{\overline{G \downarrow S}, \mathcal{A}}$ becomes satisfiable for the first time. Then, any model of that $\phi_n^{\overline{G \downarrow S}, \mathcal{A}}$ can be directly translated to an optimally reduced supervisor. We here shall remark that the minimum value of $n$ that makes $\phi_n^{\overline{G \downarrow S}, \mathcal{A}}$ satisfiable is upper bounded by the state size of integer $N$.
%\footnote{{\color{red} the state size relation with $S$?}}, since by construction we have that ${\bf OC}_{S}$ is control equivalent to $S$~\cite{B93}. The details of the encoding are now explained below. %; and each $n$-bounded finite state supervisor solution can also be translated to a partial model of the formula $\phi_n^{G \downarrow S, \mathcal{A}}$. Thus, it is possible to obtain the set of all control equivalent $n$-bounded finite state supervisors of $S$, using All-SAT solvers~\cite{GSY04},~\cite{HK12}%Consider any given instance of the $n$-bounded supervisor reduction problem with inputs being $G, \mathcal{A}$ and $M$.  $C \equiv M (\textit{mod } G)$, that is, 

Let $S'=(X', \Sigma, \xi', x_0')$ be an $n$-bounded finite state supervisor over $\mathcal{A}$, where $X':=\{x_0', x_1', \ldots, x_{n-1}'\}$ consists of $n$ states, $x_0' \in X'$ is the initial state; the partial transition function $\xi': X' \times \Sigma \longrightarrow X'$ is the only parameter that needs to be determined to ensure that $S'$ is a solution of the given instance, if a solution exists. Then, we know that $\overline{S'}$ is given by the 5-tuple $(\{x_0', x_1', \ldots, x_{n-1}', x_d'\}, \Sigma, \overline{\xi'}, x_0',\{x_0', x_1', \ldots, x_{n-1}'\})$ and we need to determine $\overline{\xi'}$. There are constraints on $\overline{\xi'}$ that can be used to reduce the search space of $\overline{S'}$. First of all, $\overline{\xi'}(x_d', \sigma)=x_d'$ for any $\sigma \in \Sigma$. Secondly, our setting assumes normality on the supervisors. For a normal supervisor $S'$ w.r.t. control constraint $(\Sigma_c,\Sigma_o)$, the observability constraint is then reduced to: for any state $x' \in X'$ and any unobservable event $\sigma \in \Sigma_{uo}$, $\xi' (x',\sigma) = x'$. This translates to:  for any state $x' \in X'$ and any unobservable event $\sigma \in \Sigma_{uo}$, $\overline{\xi'}(x',\sigma) = x'$. For convenience, we let $x_n'=x_d'$. We introduce the boolean variables $t_{x_i', \sigma, x_j'}$, where $x_i', x_j' \in X' \cup \{x_n'\}$,  and $\sigma \in \Sigma$,  in the encoding of $\overline{S'}$ with the interpretation that $t_{x_i', \sigma, x_j'}$ is true if and only if $\overline{\xi'}(x_i', \sigma)=x_j'$.

We encode the fact that $\overline{\xi'}$ is a transition function using the following constraints.
%$C$
\begin{enumerate}
%\item $t_{q_n, \sigma, q_n}$ for each $\sigma \in \Sigma$
\item $\neg t_{x_i', \sigma, x_j'} \vee \neg t_{x_i', \sigma, x_k'}$ for each $i \in [0, n-1]$, each $\sigma \in \Sigma_o$ and each $j \neq k \in [0, n]$
\item $\bigvee_{j \in [0, n]}t_{x_i', \sigma, x_j'}$ for each $i \in [0, n-1]$ and each $\sigma \in \Sigma_o$ 
\end{enumerate}
%Constraints (1) state that all the outgoing transitions from the dumb state $q_n$ are self-loops. 
We remark that Constraints (1) are imposed to ensure that $\overline{\xi'}$ is a partial function, and  Constraints (2) are imposed to ensure that $\overline{\xi'}$ is total. % and thus $C$ is complete. 
 Together, they will ensure that $\overline{\xi'}$ is a transition function and thus $\overline{S'}$ is a complete finite state automaton. Then,
\begin{center}
$\phi_n^{fsa}=\bigwedge_{i \in [0, n-1], \sigma \in \Sigma_o, j \neq k \in [0, n]} (\neg t_{x_i', \sigma, x_j'} \vee \neg t_{x_i', \sigma, x_k'}) \wedge \bigwedge_{i \in [0, n-1], \sigma \in \Sigma_o}(\bigvee_{j \in [0, n]}t_{x_i', \sigma, x_j'})$
\end{center}
shall denote the resultant formula after combining Constraints (1) and (2).

With the above constraints, we can now encode the fact that $S'$ is a finite state supervisor over $\mathcal{A}=(\Sigma_c, \Sigma_o)$ using the following extra constraints.
\begin{enumerate}
\item [3)] $\bigvee_{j \in [0, n-1]} t_{x_i', \sigma, x_j'}$ for each $i \in [0, n-1]$ and each $\sigma \in \Sigma_{uc}-\Sigma_{uo}$
\end{enumerate}
In particular, Constraint (3) is imposed to ensure controllability {\bf (C)}. We do not need to care about unobservable events $\sigma \in \Sigma_{uo}$, which surely lead to self-loops, and we do not need to care about observability, which has been taken care of. 
We shall note that the range of the index $j$ in (3), does not contain $n$, since $x_n'$ is the dump state and thus not relevant for controllability {\bf (C)}. %In normal supervisor, for any state $q_i \in Q$, unobservable event $\sigma \in \Sigma_{uo}$, there exists a self-loop transition $\overline{\delta} (q_i,\sigma)=q_i$. It is imposed to ensure observability {\bf (O)}.
%\vspace{-4pt}
%\begin{center}
%(4') $\bigwedge_{j \in [0, n-1]} (\neg t_{q_i, \sigma, q_j} \vee t_{q_i, \sigma, q_i})$ for each $i \in [0, n-1]$ and each $\sigma \in \Sigma_{uo}$
%\end{center}
%\vspace{-4pt} 
Then, 
\begin{center}
$\phi_n^{con}=\bigwedge_{i \in [0, n-1], \sigma \in \Sigma_{uc}-\Sigma_{uo}}(\bigvee_{j \in [0, n-1]} t_{x_i', \sigma, x_j'})$
\end{center}
shall denote the resultant formula ensuring controllability.

The above constraints only guarantee that $S'$ is a finite state supervisor over $\mathcal{A}$, and we still need to encode the fact that $\overline{S'}$ is a separating finite state automaton for %$(L_m(S), L_m(G) \backslash L_m(S))$ and 
$(L_m(\overline{G}) \cap L_m(\overline{S}), L_m(\overline{G}) \backslash L_m(\overline{S}))$. That is, we need to encode the fact that 
%$L_m(S) \subseteq L_m(C)$, $L_m(C) \cap (L_m(G)\backslash L_m(S))=\varnothing$,  
$L_m(\overline{G}) \cap L_m(\overline{S}) \subseteq L_m(\overline{S'})$, $L_m(\overline{S'}) \cap (L_m(\overline{G}) \backslash L_m(\overline{S}))=\varnothing$. We recall that
\vspace{-4pt}
 \begin{center}
$\overline{G \downarrow S}=(Y, \Sigma, \rho, y_0, Y_A, Y_B)$,
\end{center}
\vspace{-4pt}
where $Y_A, Y_B$ %, C_{G \downarrow S}, D_{G \downarrow S}$ 
are used to recognize the two different kinds of languages %$L_m(S)$, $L_m(G) \backslash L_m(S), 
$L_m(\overline{G}) \cap L_m(\overline{S})$ and $L_m(\overline{G}) \backslash L_m(\overline{S})$, respectively. The remaining constraints are formulated based on $\overline{S'}$ and $\overline{G \downarrow S}$. 

%the two basic properties in verifying separating partial DFAs, that is, 
Language inclusion and emptiness of language intersection can be checked by using the  synchronous product operation, we only need to track the reachable states in the synchronous product of $\overline{S'}$ and $\overline{G \downarrow S}$. We now introduce, as in~\cite{D12}, auxiliary Boolean variables $r_{x', y}$, where $x' \in X' \cup \{x_n'\}$ and $y \in Y$, with the interpretation that if state $(x', y)$ is reachable from the initial state $(x_0', y_0)$ in the synchronization of $\overline{S'}$ and $\overline{G \downarrow S}$, then $r_{x', y}$ is True. We have the following constraints.

%Boolean variables $x_{q_i, q'}$ and  $t_{q_i,\sigma,q''}$ are true when state ($q_i,q'$) is reachable from initial state ($q_0,q_{G\downarrow S,0}$) and $\bar\delta(q_i,\sigma)=q_j$. Also, $q'\in Q_{G\downarrow S}$, we have $\delta_{G\downarrow S}(q',\sigma)=q''$ meaning that $q''\in Q_{G\downarrow S}$. Since $q_i$ is reachable from $q_o$ and there exist $\sigma$ transition from $q_i$ to $q_j$, $q_j$ is reachable from $q_o$ as well. Therefore, We can know that the state ($q_j,q''$) is reachable from ($q_0,q_{G\downarrow S,0}$) and $x_{q_j,q''}$  is true. If $q'\in L(S)$ and $q'\in L(S/G)$.

\begin{enumerate}
\item [5)] $r_{x_0', y_0}$%, where $q_0^G, q_0^C, q_0^M$ are respectively the initial states of $G, C, M$
\item [6)] $r_{x_i', y_1} \wedge t_{x_i', \sigma, x_j'} \Longrightarrow r_{x_j', y_2}$, for each $i, j \in [0, n]$, each $y_1 \in Y$, each $\sigma \in \Sigma$ and for each $y_2 \in Y$ such that $y_2=\rho(y_1, \sigma)$
\item [7)] $\neg r_{x_n', y}$, for each $y \in Y_A$
\item [8)] $\bigwedge_{i \in [0, n-1]} \neg r_{x_i', y}$, for each $y \in Y_B$
\end{enumerate}
In particular, Constraints (5) and (6) are used to propogate the constraints on $r_{x', y}$, based on the  synchronous product construction and the inductive definition of reachability. Based on Constraints (5) and (6),  Constraints (7) are used to ensure $L_m(\overline{G}) \cap L_m(\overline{S}) \subseteq L_m(\overline{S'})$; Constraints (8) are used to ensure $L_m(\overline{S'}) \cap (L_m(\overline{G}) \backslash L_m(\overline{S}))=\varnothing$.

Let the conjunction of Constraints (5), (6), (7) and (8) be denoted by formula  $\phi_n^{sep}$, which enforces the separating finite state automaton constraint. We have 
\begin{center}
$\phi_n^{sep}=r_{x_0', y_0} \wedge \bigwedge_{i, j \in [0, n], y_1 \in Y, \sigma \in \Sigma, y_2=\rho(y_1,\sigma)}(\neg r_{x_i', y_1} \vee \neg t_{x_i', \sigma, x_j'} \vee r_{x_j', y_2}) \wedge \bigwedge_{y \in Y_A}\neg  r_{x_n', y} \wedge \bigwedge_{y \in Y_B, i \in [0, n-1]}\neg r_{x_i', y}$,
\end{center}
which has been converted to CNF.

Let $\phi_n^{\overline{G\downarrow S},\mathcal{A}}:= \phi_n^{dfa}\wedge\phi_n^{con} \wedge \phi_n^{sep}$. It is clear that $\phi_n^{\overline{G\downarrow S},\mathcal{A}}$ is in CNF.  We note that Constraints (3) implies Constraints (2) on the part where $\sigma \in \Sigma_o \cap \Sigma_{uc}$, making Constraint (2) partly redundant. By solving $\phi_n^{\overline{G\downarrow S},\mathcal{A}}$ using a SAT solver~\cite{what12}, we can get $\overline{S'}$ by interpreting the Boolean variables $t_{x_i', \sigma, x_j'}$ on $\overline{\xi'}$, which can be used to recover $S'$.
We have the following result.
%According to the corollary 1, since the synchronous product of $S'$ and $G\downarrow S$ can separate two disjoint language $L(S)\cap L(G)$ and $L(G)\backslash L(S)$, implying $\overline{S'}$ as the complement automaton. $\phi_n^{fsa}$ is true means that $\overline{\delta}$ is a total function. $\phi_n^{co}$ are supposed to be satisfiable guarantees control constraints $\mathcal{A}$ controllablity and observablity.

\begin{theorem}
The formula $\phi_n^{\overline{G\downarrow S},\mathcal{A}}$ is correct. That is, there exists an $n$-bounded finite state supervisor $S'$ that solves the instance with plant $G$, supervisor $S$ and control constraint $\mathcal{A}$  if and only if $\phi_n^{\overline{G \downarrow S}, \mathcal{A}}$ is satisfiable.
\end{theorem}

After obtaining $S'$, we shall perform the non-attackability verification. This is explained in the next subsection (see~\cite{Lin2018} for more details).

%\item [7)] $x_{q_i, q'} \Longrightarrow m_{q_i}$, for each $i \in [0, n-1]$ and each $q' \in A_{G \downarrow S}$
%\item [8)] $x_{q_i, q'} \Longrightarrow \neg m_{q_i}$, for each $i \in [0, n-1]$ and each $q' \in B_{G \downarrow S}$
%\item [9)] $\neg x_{q_n, q'}$, for each $q' \in C_{G \downarrow S}$
%\item [10)] $\bigwedge_{i \in [0, n-1]} \neg x_{q_i, q'}$, for each $q' \in D_{G \downarrow S}$

{\bf Attackability Verification}:
We here recall the steps used in the algorithm for verifying non-attackability, i.e., Algorithm 3.

{\bf 1. Annotation of the Supervisor}

Given the supervisor $S=(X, \Sigma, \zeta, x_0)$, we produce the annotated supervisor 
\begin{center}
$S^{A}=(X, \Sigma_o \times \Gamma \cup \Sigma_{uo}, \zeta^{A}, x_0)$, 
\end{center}
where $\zeta^{A}: X \times (\Sigma_o \times \Gamma \cup \Sigma_{uo}) \longrightarrow X$ is the partial transition function, which is defined as follows:
\begin{enumerate}
\item For any $x, x' \in X$, $\sigma \in \Sigma_o$, $\gamma\in \Gamma$, $\zeta_A(x, (\sigma, \gamma))=x'$ iff $\zeta(x, \sigma)=x'$ and $\gamma=\{\sigma \in \Sigma \mid \zeta(x', \sigma)!\}$.
\item For any $x, x' \in X$, $\sigma \in \Sigma_{uo}$, $\zeta_A(x, \sigma)=x'$ iff $\zeta(x, \sigma)=x'$.
\end{enumerate}

%We need to transfer the annotation in $P(S^A)$ to $S \lVert G$ to obtain a transducer structure that maps string executions in $L(S \lVert G)$ to attacker's observation sequences. This is not that difficult as we only need to synchronize the plant $G$ and the annotated supervisor $S^A$ using a dedicated synchronous product operation, which transfers the attacker's observation annotation, which then encodes the function $\hat{P}_{o, A}^{V} \circ P_o: L(S \lVert G) \rightarrow ((\Sigma_{o, A} \cup \{\epsilon\}) \times \Gamma)^*$. Based on the product of $G$ and $S^A$, we can compute every set of strings in $L(S \lVert G)$ that can be mapped to the same attacker's observation sequence, via a subset construction w.r.t. the attacker's observation alphabet. To determine whether an attack shall be established for an attacker's observation sequence, we need to synchronize $G$ and $S^A$ with $H$ before the determinization is performed. The overall product operation will be referred to as the {\em generalized synchronous product} operation, which is explained below.

{\bf 2. Generalized Synchronous Product:}

Given the plant $G=(Q, \Sigma, \delta, q_0)$, the annotated supervisor $S^{A}=(X, \Sigma_o \times \Gamma \cup \Sigma_{uo}, \zeta^{A}, x_0)$ and the damage automaton $H=(Z, \Sigma, \eta, z_0)$, the generalized synchronous product $GP(G, S^A, H)$ of $G$, $S^{A}$ and $H$ is given by
\begin{center}
$(Q \times X \times Z \cup \{\bot, \top\}, \Sigma^{GP}, \delta^{GP}, (q_0, x_0, z_0))$,
\end{center}
Where, the state $\bot$ indicates a failed attack, while the state $\top$ indicates a successful attack. $\bot, \top$ are two new states that are different from the states in $Q \times X \times Z$, $\Sigma^{GP}=\Sigma_o \times ((\Sigma_{o, A} \cup \{\epsilon\}) \times \Gamma) \cup \Sigma_{uo} \times \{\epsilon\} \cup \Sigma_{c, A}$ and the partial transition function 
\begin{center}
$\delta^{GP}: (Q \times X \times Z \cup \{\bot, \top\}) \times \Sigma^{GP} \longrightarrow Q \times X \times Z \cup \{\bot, \top\}$
\end{center}
is defined as follows.
\begin{enumerate}
\item For any $q, q' \in Q$, $x, x' \in X$, $z, z' \in Z$, $\gamma \in \Gamma$, $\sigma \in \Sigma_{o, A}$, $\delta^{GP}((q, x, z), (\sigma, (\sigma, \gamma)))=(q', x', z')$ iff $\delta(q, \sigma)=q'$, $\zeta^A(x, (\sigma, \gamma))=x'$ and $\eta(z, \sigma)=z'$.
\item For any $q, q' \in Q$, $x, x' \in X$, $z, z' \in Z$, $\gamma \in \Gamma$, $\sigma \in \Sigma_{o}-\Sigma_{o, A}$, $\delta^{GP}((q, x, z), (\sigma, (\epsilon, \gamma)))=(q', x', z')$ iff $\delta(q, \sigma)=q'$, $\zeta^A(x, (\sigma, \gamma))=x'$ and $\eta(z, \sigma)=z'$.
\item For any $q, q' \in Q$, $x, x' \in X$, $z, z' \in Z$, $\sigma \in \Sigma_{uo}$, $\delta^{GP}((q, x, z), (\sigma, \epsilon))=(q', x', z')$ iff $\delta(q, \sigma)=q'$, $\zeta^A(x, \sigma)=x'$ and $\eta(z, \sigma)=z'$.
\item For any $q \in Q$, $x \in X$, $z\in Z$, $\sigma \in \Sigma_{c, A}$, $\delta^{GP}((q, x, z), \sigma)=\top$ iff $\delta(q, \sigma)!$, $\neg \zeta(x, \sigma)!$ and $\eta(z, \sigma) \in Z_m$.
\item For any $q \in Q$, $x \in X$, $z\in Z$, $\sigma \in \Sigma_{c, A}$, $\delta^{GP}((q, x, z), \sigma)=\bot$ iff $\delta(q, \sigma)!$, $\neg \zeta(x, \sigma)!$ and $\eta(z, \sigma) \notin Z_m$.
\end{enumerate}

{\bf 3. Subset Construction w.r.t. the Attacker's Observation Alphabet}:

Given $GP(G, S^A, H)$, we shall now perform a subset construction on $GP(G, S^A, H)$, with respect to the attacker's observation alphabet. 

Let $GPS(G, S^A, H)$ denote the sub-automaton of $GP(G, S^A, H)$ with state space restricted to $Q \times X \times Z$. The alphabet of $GPS(G, S^A, H)$ is $\Sigma^{GPS}=\Sigma_o \times ((\Sigma_{o, A} \cup \{\epsilon\}) \times \Gamma) \cup \Sigma_{uo} \times \{\epsilon\}$.
Let $GPS^2(G, S^A, H)$ denote the automaton that is obtained from $GPS(G, S^A, H)$ by projecting out the  first component of the event of each transition. Let $SUB(GP(G, S^A, H))$ denote the the automaton that is obtained from $GPS^2(G, S^A, H)$ by determinization. In addition, we augment $SUB(GP(G, S^A, H))$ with a labeling function $Lf: Y \mapsto 2^{\Sigma_{c, A}}$, which is defined such that for any $y \in Y$ and any $\sigma \in \Sigma_{c, A}$, $\sigma \in Lf(y)$ iff there exists some $v \in y$ such that, 1) $\delta^{GP}(v, \sigma)=\top$ and 2) for any $v' \in y$, $v' \neq v$ implies $\delta^{GP}(v', \sigma) \neq\bot$. %{\color{red} Then we augment $SUB(GP(G, S'^A, H))$ with a labeling function $Lf: Y \mapsto 2^{\Sigma_{c, A}}$, which is defined such that for any $y \in Y$ and any $\sigma \in \Sigma_{c, A}$, $\sigma \in Lf(y)$ iff for all states in $v \in y$ such that, $\delta^{GP}(v, \sigma)\neq\top$}.

%{\color{red} it is nice to add a final theorem to say that the overall algorithm is correct..If it terminates, the computed supervisor is resilient, behavior preserving and of minimum state size}

We here recall the following result~\cite{Lin2018}, which states that Algorithm \textbf{NA}($G,S',H$) is correct.
\begin{theorem}
Let the plant $G$ and the normal supervisor $S$ be given. Then, $(G, S)$ is unattackable with respect to $(\Sigma_{c, A}, \Sigma_{o, A})$ and $L_{dmg}$ iff for every reachable state $y$ in $SUB(GP(G, S^A, H))$ it holds that $Lf(y) = \varnothing$.
\end{theorem}

\section{Conclusion}
In this paper, we have proposed and addressed the problem of supervisor obfuscation in actuator enablement attack scenario, in the setting where the attacker is able to eavesdrop the control commands issued by the supervisor and under a normality assumption on the actuator attackers and supervisors. 

The algorithm proposed in this paper for solving the supervisor obfuscation problem relies on an SAT solver for computing behavior-preserving supervisors; a behavior-preserving supervisor is accepted as a solution of the problem only when it passes the non-attackability verification. The method used in this paper enumerates each possible behavior-preserving supervisor and then verifies its non-attackability. Although SAT solver is relatively efficient in computing behavior-preserving supervisors, the enumeration process is still causing massive degradation on the performance of the overall algorithm. In future work, we intend to develop more efficient algorithms that can be used for directly searching for obfuscated supervisors.  %We can see the weakness in the exponential time and without a specified condition to terminate the process. We intend to improve SAT algorithm, characterizing condition ....{\color{red} to continue}
%\item encode the supervisor reduction problem into SMT formulas and solve it more effectively.  
%\end{enumerate}

%{\color{red} in conclusion, usually no math symbol is used as in the abstract..only a general discussion is needed. In particular, if limitation and future work has been written, it is likely a solid conclusion section.}

%{\color{red} I can take care of conclusion section}

%In this paper, We provide an algorithm to compute the upper $n$-bounded obfuscated supervisor under actuator enablement attack within discrete event systems formalism. At first, there exists successful actuator  enablement attacker on $(G,S)$ w.r.t $\Sigma_{c,A},\Sigma_{o,A}$ and $L_{dmg}$. By replacing the original supervisor $S$ with the new obfuscated supervisor $S'$, which ensures the same closed-loop behavior, observalibity and controllability, then $(G,S')$ turns to be unattackable w.r.t. $(\Sigma_{c,A},\Sigma_{o,A})$ and damage-inflicting automaton $H$. 
%The process to obtain supervisor $S'$ can be regarded as supervisor reduction problem, which is similar with minimal separating finite state automaton w.r.t control constraint and close-loop behaviour solved by SAT approach. 
%{\color{red}Should I explain the difference between the normal supervisor reduction(not in polynomial time) and with the use of SAT solver}
%Therefore, this reduces to an NP-hardness problem.

\end{document}